\documentclass[prd,abs,superscriptaddress,nofootinbib,eqsecnum,twocolumn]{revtex4-1}

\pdfoutput=1

\usepackage{graphicx}  
\usepackage{dcolumn}   
\usepackage{amssymb}   
\usepackage{xcolor}
\usepackage{graphics} 
\usepackage{hyperref}
\usepackage{amsfonts}
\usepackage{latexsym}
\usepackage{slashed}
\usepackage{bm}  
\usepackage[normalem]{ulem} 
\usepackage{amsmath,mathtools} 

\usepackage{array}
\usepackage{wrapfig}
\usepackage{multirow}
\usepackage{tabularx}

\usepackage{float}

\begin{document}

\title{First order phase transitions within Weyl type of materials at low temperatures}

\author{Y. M. P. Gomes} \email{yurimullergomes@gmail.com}
\affiliation{Departamento de F\'{\i}sica Te\'{o}rica, Universidade do Estado do Rio de Janeiro, Rio de Janeiro, RJ 20550-013, Brazil}

\author{Everlyn Martins} \email{everlyn.martins@posgrad.ufsc.br}
\affiliation{Departamento de F\'{\i}sica, Universidade Federal de Santa
  Catarina, Florian\'{o}polis, SC 88040-900, Brazil}
 
\author{Marcus Benghi Pinto} \email{marcus.benghi@ufsc.br}
\affiliation{Departamento de F\'{\i}sica, Universidade Federal de Santa
  Catarina, Florian\'{o}polis, SC 88040-900, Brazil}
  
\author{Rudnei O. Ramos} \email{rudnei@uerj.br}  
\affiliation{Departamento de F\'{\i}sica Te\'{o}rica, Universidade do Estado do Rio de Janeiro, Rio de Janeiro, RJ 20550-013, Brazil}
\affiliation{Physics Department, McGill University, Montreal, Quebec, H3A 2T8, Canada}

\begin{abstract} 

We analyze the possible dynamical chiral symmetry breaking patterns
taking place within  Weyl type of materials. Here, these systems are
modeled by the (2+1)-dimensional Gross-Neveu model with a tilt in the
Dirac cone. The optimized perturbation theory (OPT) is employed in
order to evaluate the effective  potential at finite temperatures and
chemical potentials beyond the traditional large-$N$ limit. The nonperturbative
finite-$N$ corrections generated by the OPT method and its associated
variational procedure show that a first-order phase transition boundary,
missed at large $N$, exists in the regime of low temperatures and
large chemical potentials. This result, which represents our main
finding, implies that one should hit a region of mixed phases when
exploring the low-temperature range. The associated first order
transition line, which starts at $T=0$, terminates at a tricritical
point such that the transitions taking place at high $T$ are of the
second kind. In particular, we discuss how the tilt in the Dirac cone
affects the position of the tricritical point as well as the values of
critical temperature and coexistence chemical potential among other
quantities. Some experimental implications  and  predictions are also
briefly discussed.  

\end{abstract}

\maketitle 
 
\section{Introduction}
 
Dirac materials are condensed matter systems whose  excitations can be
described by a Dirac-type equation, i.e., they obey dispersion
relations linear in  momentum.  As a prime example, one can cite
graphene~\cite{Novoselov:2005kj}, which represents the first material
discovered with this property. The same linear property can be seen in
three- and two-dimensional Weyl
semimetals~\cite{Grushin:2012mt,tchoumakov2016,soluyanov2015,wan2011,goerbig2008},
and on the metallic edge of topological
insulators~\cite{Hasan:2010hm}. As a result, the electronic
quasi-particles in all the aforementioned systems have linear
dispersion relations like a Dirac fermion.
 
 A more complex structure can be seen in the three-dimensional Weyl
 semimetals (3DWSM), where the linear dispersion is combined with
 chiral symmetry to ensure topological protection against gap
 formation.  Since we do not expect  solid-state materials to exactly
 obey the stringent Lorentz invariance~\cite{Kostelecky:2021bsb}, this
 opens a new range of possible phenomena, among them, the tilting of
 the Dirac cone.  There is a possibility to form Weyl excitations in
 two dimensions as well, and the resulting dynamics show that in this
 case the Dirac cone becomes tilted, and, consequently, Lorentz
 invariance is explicitly broken in these systems.  However, the
 topological properties of the two-dimensional Weyl semimetals (2DWSM)
 are less restrictive to gap formation and, thus, can be sensitive to
 doping, possibly favoring its application. In addition,  the study of
 phase transitions in planar systems, e.g., like  graphene and
 similar type of systems, has become a timely problem. This motivates
 us to consider the 2DWSM case in this work.

Here, the tilted 2DWSM system will be described by the   four-fermion
Gross-Neveu (GN) model~\cite{Gross:1974jv} in $(2+1)$ dimensions. One
appealing feature of this model is that it allows  some condensed
matter systems, for which chiral symmetry plays a major role, to  be
treated within the framework of powerful quantum field theory
techniques. In this vein, several applications considered the GN   in
order to seek knowledge about the gap formation in both
linear~\cite{Caldas:2008zz} and planar  condensed matter
systems~\cite{Caldas:2009zz,Ramos:2013aia,Zhukovsky:2017hzo,Ebert:2015hva,Klimenko:2013gua,Klimenko:2012qi,Khunjua:2021fus}.
As far as  2DWSM tilted systems are concerned,  previous large-$N$
(LN) applications~\cite{Gomes:2021nem,Gomes:2022dmf} indicated that
the gap formation occurs  at lower temperatures and/or chemical
potentials  in comparison with what happens within  untilted systems,
like graphene.  In other words, the presence of a tilt in the Dirac
cone disfavors chiral symmetry breaking (CSB) so that the
corresponding region  of the phase diagram shrinks when considering a
nonnull tilting parameter which represents the magnitude of the tilt
vector ${\bf t}$.   Another important point to be addressed
regards the order of the phase transitions taking place at the phase
diagram boundaries.  The seminal LN applications to the untilted GN
model in
$(2+1)-$dimensions~\cite{Klimenko:1987gi,Rosenstein:1988dj,Rosenstein:1988pt}
have shown  that the phase transition is of the second kind in all
regions of the phase diagram except at (exactly) $T=0$, where it
happens to be of the first kind. Using the LN approximation, some of
the present authors have shown~\cite{Gomes:2021nem}  that the presence
of a tilt does not alter this picture in a qualitative way. It is an
important issue to verify whether this phase transition behavior would
persist or change when going beyond the LN approximation. As a matter
of fact, we can rightfully criticize the reliability of the LN
approximation, especially since real physical systems,  like  Weyl and
Dirac materials, only have a finite number of fermion nodes, $N\sim
{\cal O}(1)$, which can indicate that the LN approximation, when
applied to these systems, can be potentially inaccurate.  In fact, in
the absence of tilting, i.e., ${\bf t}=0$, the somewhat unexpected  phase
portrait obtained for the GN model has been challenged by Kogut and
Strouthos~\cite{Kogut:1999um}, who used continuity arguments in order
to argue that the first order transition (occurring at $T=0$) should
persist at  finite temperatures, terminating at a tricritical point,
where a second order transition line should start. Using lattice
Monte Carlo simulations, these authors  have studied the untilted
planar GN model at finite $N$ and have predicted that a tricritical
point should indeed exist at  finite (low) values of $T$ and large
values of the chemical potential. However,  within the numerical
precision of their simulations, they were unable to give its exact
location\footnote{Earlier lattice simulations at finite $N$  have also
found some evidence for a tricritical point at non-vanishing
temperatures~\cite{Hands:1992be,Hands:1992ck}.}. Motivated by those
results, some of the present authors have used the {\it optimized
  perturbation theory} (OPT) \cite{ana,moshe}  in order to investigate
how finite-$N$ corrections could affect the phase diagram boundaries
predicted at LN in the untilted
case~\cite{Kneur:2007vj,Kneur:2007vm}. The results obtained from such
application have confirmed that, when $N$ is finite, a first-order
transition line (starting at $T=0$) terminates at a tricritical point
in the low-$T$ and high-$\mu$ region of the phase diagram. It has also
been observed that the length of this transition line increases with
decreasing $N$. Later, those applications have been extended to  study
the phase diagram in the presence of a magnetic field \cite
{Kneur:2013cva}. It is worth mentioning that some of the OPT
predictions to the GN model in $2+1\,d$ performed in
Refs. \cite{Kneur:2007vj,Kneur:2007vm, Kneur:2013cva}  have been
confirmed by recent lattice simulations at finite $N$ \cite
{lenz1,lenz2}.

The aim of this work is to extend the OPT application performed
in Refs.~\cite{Kneur:2007vj,Kneur:2007vm} to the case when Lorentz
symmetry is broken through the tilting of the Dirac cone, which is
relevant to understand Weyl-type real planar condensed matter
systems. In this way, we will be able to go beyond the LN results
obtained in Ref.~\cite {Gomes:2021nem} and then gauge the importance
of the finite $N$ corrections, which the nonperturbative OPT method
brings,  in leading to a more reliable phase diagram portrait for
${\bf t}\neq 0$. This is of particular importance when one wishes to model
real physical systems through quantum field theory methods.  We can
anticipate here that, backed by those previous applications going
beyond the LN approximation, a tricritical point will also show up at
${\bf t}\ne 0$, when $N$ is finite. Then, we will be able to investigate how
the tilt influences different quantities, such as the length of the
first-order transition line and coexistence region, as well as the
location of the tricritical point among other relevant quantities. 
 
The remainder of this work is organized as follows.  In
Sec.~\ref{section2}, we show the main aspects of the low-energy model
describing two-dimensional Weyl materials. The OPT method is described
in Sec.~\ref{section3} and then used in order to evaluate the finite
$N$ corrections to the effective  potential. In Sec.~\ref{section4},
we discuss the optimization process required by the OPT method. The
main results are obtained in Sec.~\ref{section5}, while the
conclusions are presented in  Sec.~\ref{section6}.  Throughout this
paper, we  consider natural units by setting $k_B =\hbar = c \equiv
1$. 
 
\section{The model}
\label{section2}

The low-energy Hamiltonian describing a system of  quasi-particles
with a tilted Dirac cone can be written
as~\cite{Gomes:2021nem,Gomes:2022dmf}
\begin{equation}\label{h1}
H(p) = v_F [ ({\bf t} \cdot {\bf p})\tau^0 + (\xi_x p_x) \tau^x +
  (\xi_y p_y) \tau^y] ,
\end{equation}
where $v_F$ is the {}Fermi velocity, ${\bf t}$ is the tilt vector,
$\xi_{x,y}$ represent  anisotropy factors, while $\tau^0 = I_{2 \times 2}$
and $\tau^{x,y}$ are the Pauli matrices. When $|{\bf t}|<1$, we have a
type-I Weyl semimetal, and when $|{\bf t}|>1$ the semimetal is of
type-II.  {}From Eq.~\eqref{h1} we obtain the energy spectrum as given
by
\begin{equation}
    E_\epsilon({\bf p}) = v_F \left[ {\bf t} \cdot {\bf p} + \epsilon
      \sqrt{(\xi_x p_x)^2+(\xi_y p_y)^2} \right]\,,
\end{equation}
with $\epsilon =\pm 1$ for the conduction and valence
bands. {}Furthermore, in order to ensure that $\epsilon = +1$ and
$-1$ are correctly associated with positive and negative
energy states, the effective tilt parameter $|{\bf \tilde{t}}|$, defined as
\begin{equation} 
|{\bf \tilde{t}}|\equiv  \sqrt{\left(
  \frac{t_x}{\xi_x}\right)^2 +\left(\frac{t_y}{\xi_y}\right)^2} ,
\label{ttilde}
\end{equation}
must satisfy $|{\bf \tilde{t}}| < 1$.

Note that Eq.~(\ref{h1}) gives the general Hamiltonian for a generic Weyl-type fermionic two-dimensional system. In order to embrace all the degrees of freedom of the fermionic quasi-particles one can construct a four-component spinor, which incorporates both chiralities. In particular note that the breaking of chiral symmetry  introduces a mass term that washes out the Weyl separation between distinct chiralities. Therefore, the quasi-particles in the chiral broken phase are Dirac type, meaning that the corresponding system represents a Dirac material. Only when the chiral symmetry has been restored, and the chiralities decouple,  one has a Weyl-type type of material.
The presence of both chiralities is essential for the appearance  of a non-vanishing mass gap, which in the present case is generated by the dynamical breaking of chiral symmetry (in particular note that the Nielsen-Ninomiya no-go theorem~\cite{Nielsen:1980rz,Nielsen:1981xu} is not violated in our model). 
Therefore, one can build a Lagrangian density
for the four-component spinor $\psi$ and which also include the tilting of the Dirac cone as~\cite{Gomes:2021nem}
\begin{equation}\label{lagr1}
\mathcal{L}=\sum_k\bar{\psi}_k\left( M^{\mu \nu}\gamma_\mu
\partial_\nu\right) \psi_k ,
\end{equation}
where the matrix $M$ is given by
\begin{equation} 
M = \begin{pmatrix} 1 & - v_F t_x & -v_F t_y\\ 0 & -v_F \xi_x & 0\\ 0
  & 0 & -v_F\xi_y 
\end{pmatrix},
\end{equation}
while the Dirac gamma matrices are defined as
\begin{equation}
\gamma^\mu = \tau^\mu \otimes \begin{pmatrix} 1 & 0 \\ 0 & -1
\end{pmatrix},
\end{equation}
with $\tau^\mu = (\tau_z, i \tau_x , i \tau_y)$, $\bar{\psi} =
\psi^\dagger \gamma^0$ where $\tau_z$ represents the third Pauli
matrix. The $\gamma$-matrices respect the algebra $\{\gamma^\mu,
\gamma^\nu \}= 2 \eta^{\mu \nu}$, with $\eta^{\mu \nu}= diag
(+,-,-)$. 

Based on Eq.~\eqref{lagr1} one can construct a GN-like model that
describes the excitonic pairing in the 2DWSM. This excitonic
condensate will break the chiral symmetry creating a gap between
valence and conducting bands. In terms of a local excitonic
self-interaction, the system can be described by the following GN-like
Lagrangian density (in Euclidean space)\footnote{Note that in all of
our expressions, $N$ only counts the effective number of fermion
fields (e.g., the number of bands), while the spin degrees of freedom
are explicitly accounted for when taking the trace in all quantum and
thermal corrections to be evaluated for the free energy.},
\begin{equation}
\mathcal{L} = \bar{\psi}_k\left( M^{\mu \nu}\gamma_\mu
\partial_\nu\right) \psi_k -  \frac{\lambda v_F}{2 N} \left (
\bar{\psi}_k \psi_k \right)^2,
\label{GNlag}
\end{equation}
where a sum over $k=1,\ldots,N$ is implied.

One should notice that local four-fermion interactions like in Eq.~(\ref{GNlag}) are well motivated. These interactions can describe for example the effective
interaction between electron and phonons in materials~\cite{phonons},
or also the effects of impurity and disorder~\cite{Zhao:2018jro,Wang:2018trs}.
The same type of local four-fermion interaction of the type of the GN model has also been extensively used in the case of honeycomb
lattice searching for gap generation~\cite{Drut:2009aj,Herbut:2006cs,Son:2007ja,quant1,Drut:2008rg,Juricic:2009px}.
Contact four-fermion interaction terms are also expected to appear
in the continuum limit derived from the original lattice tight-binding
model~\cite{tb1,Herbut:2006cs,tb2,tb3,Aleiner:2007va}
and they appear in addition to the Coulomb
interaction.

\section{The OPT free energy}
\label{section3}

Let us briefly describe in this section the OPT method and its
implementation.  The general procedure through which the OPT is
implemented requires modifying the Lagrangian density of the model
such that (see Refs.~\cite{Kneur:2007vj,Kneur:2007vm} and also, for a
recent review, Ref.~\cite{Yukalov:2019nhu} and references there in)
\begin{equation}
  \mathcal{L} \rightarrow\mathcal{L}^{\delta} =
  (1-\delta)\mathcal{L}_{0}(\eta) + \delta \mathcal{L} , 
\label{inter}
\end{equation}
where, in the above equation, $\mathcal{L}_{0}$ stands for the
Lagrangian density of the free theory, which is modified by an
arbitrary mass parameter, $\eta$. The other quantity, $\delta$,  makes
the role of a bookkeeping parameter,  allowing for a perturbative
expansion to a given order, $\delta^n$. As one can see,
Eq.~(\ref{inter}) allows for an  interpolation between the free
(soluble) theory, at $\delta=0$, and the (originally) interacting
one, at $\delta=1$.  For the particular model represented by the
Lagrangian density (\ref{GNlag}) this implementation can then be
described as follows.  We start by redefining the coupling constant
term according to $\lambda \to \delta \lambda$, while also adding the
Gaussian term $(1-\delta)\eta\, \bar{\psi}_k \psi_k$ to
Eq.~(\ref{GNlag}). Here, during the evaluations, the bookkeeping
parameter $\delta$ is formally treated as being $\delta \ll 1$ such
that a given physical quantity, $\Phi$, can be evaluated as a
perturbative series in powers of $\delta$. At the end, one restores
this parameter to its original value by setting $\delta=1$. The
Gaussian term contains a Lagrangian multiplier represented by the
arbitrary mass parameter $\eta$, which can be optimally fixed by
requiring that the relevant physical quantity  being computed
satisfies a variational requirement. In most  applications involving
the OPT, the  principle of minimal sensitivity
(PMS)~\cite{Stevenson:1981vj} is considered. This criterion consists
in applying the following variational condition to the physical
quantity under consideration
\begin{equation}
\frac{ d \Phi}{d \eta} \Big{|}_{\bar{\eta}, \delta=1} = 0\,.
\label{pms}
\end{equation}
As one can easily check, the deformed OPT Lagrangian density generated
from Eq.~(\ref{GNlag}) and given by
\begin{equation}
\mathcal{L} = \bar{\psi}_k\left( M^{\mu \nu}\gamma_\mu
\partial_\nu\right) \psi_k + (1-\delta) \eta \bar{\psi}_k \psi_k  -
\delta \frac{\lambda v_F}{2 N} \left ( \bar{\psi}_k \psi_k \right)^2,
\label{interGN}
\end{equation}
correctly implements the general interpolation prescription defined by
Eq.~(\ref{inter}).  Here, the OPT will be used to evaluate the
so-called effective potential, $V_{\rm eff}(\sigma_c)$, which in quantum
field theories can generate all one-particle irreducible Green's
functions with zero external momentum. Usually, the effective
potential incorporates radiative (quantum) corrections to the
classical potential representing a scalar (classical) field,
$\sigma_c$ \cite {ryder}. Therefore,  $V_{\rm eff}(\sigma_c)$ is of utmost
importance  for studies related to dynamical symmetry breaking. In
statistical mechanics the analog of $V_{\rm eff}(\sigma_c)$ is Landau's
free energy density so that the pressure and the thermodynamic
potential can be obtained by minimizing $V_{\rm eff}(\sigma_c)$ quantity
with respect to $\sigma_c$. One then obtains, $P= - \Omega = -
V_{\rm eff}({\bar \sigma}_c)$ where ${\bar \sigma}_c=\langle \sigma
\rangle_0$ represents an order parameter (see Refs. \cite
       {ryder,lebellac} for further details).

As usual, the evaluation of the effective  potential  can be
facilitated by bosonizing the original four-fermion theory through a
Hubbard-Stratonovich transformation. In this case, an auxiliary
bosonic field ($\sigma$) can be introduced when the following
quadratic term 
\begin{equation}
\frac{\delta N}{2 \lambda} \left(\sigma +\frac{\lambda}{N}
\bar{\psi}_k \psi_k  \right )^2,
\end{equation}
is added to Eq.~(\ref{interGN}. As one can check, the Euler-Lagrange
equations for $\sigma$ lead to $\sigma = - \lambda/N \bar{\psi}_k
\psi_k  $ Then, the original four-fermion vertices are replaced by
Yukawa vertices yielding
\begin{equation}
\mathcal{L} = \bar{\psi}_k\left( M^{\mu \nu}\gamma_\mu \partial_\nu-
\hat{\eta}\right) \psi_k +  \frac{\delta N}{2 \lambda v_F} \sigma^2,
\end{equation}
where $\hat{\eta} = \eta - \delta \left(\eta-\sigma \right)$. Then,
following Refs. ~\cite{Kneur:2007vj,Kneur:2007vm}, one can obtain the
order-$\delta$ result\footnote{In the following, to make the notation
less cumbersome,  we shall denote the classical field, $\sigma_c$,
simply by $\sigma$ hoping that this slight abuse of notation will not
cause further confusion.}
\begin{eqnarray}\nonumber
\frac{V_{\rm eff}}{N} &=&\frac{\delta}{2 \lambda v_F} \sigma^2 \\\nonumber
&+& i \int \frac{d^3 p}{(2\pi)^3} {\rm tr} \ln \left( M^{\mu
  \nu}\gamma_\mu p_\nu - \eta\right)    \\\nonumber &+&  \delta i \int
\frac{d^3 p}{(2\pi)^3} {\rm tr}\frac{(\eta-\sigma)}{M^{\mu
    \nu}\gamma_\mu p_\nu - \eta + i \epsilon}\\\nonumber &-& \delta
\frac{i}{2 }  \int \frac{d^3 p}{(2\pi)^3} {\rm tr}
\frac{\Sigma_{\rm exc}(\eta)}{M^{\mu \nu}\gamma_\mu p_\nu - \eta + i
  \epsilon} \,,\\
\label{OPTlag}
\end{eqnarray}
where
\begin{equation}
\Sigma_{\rm exc}(\eta) =  -\frac{i  \lambda v_F}{ N} \int \frac{d^3 q
}{(2\pi)^d} \frac{1}{M^{\mu \nu}\gamma_\mu q_\nu - \eta + i \epsilon}
,
\end{equation}
represents the leading-order exchange contribution to the self-energy.
Thermodynamic properties can be conveniently described by adopting
the Matsubara formalism, where the integrals over momentum in
Eq.~(\ref{OPTlag}) are replaced by
\begin{equation}
 \int \frac{d^3p}{(2\pi)^3} \to T \sum_{j=-\infty}^{+\infty}  \int
 \frac{d^{2}p}{(2\pi)^{2}}~~,
\end{equation} 
and $p_0 \to i\omega_j + \mu$, where $\omega_j =(2 j - 1) \pi  T$ are
the usual Matsubara's frequencies for fermions. By explicitly
performing the sum over the Matsubara's frequencies  in
Eq.~(\ref{OPTlag}), one obtains 
\begin{eqnarray}\nonumber
\frac{V_{\rm eff}(\sigma,\eta)}{N} &=& \frac{\delta}{2 \lambda v_F}
\sigma^2 -    \frac{2}{\xi_x \xi_yv_F^2} I_0+ 4 \delta
\frac{(\eta-\sigma)\eta}{\xi_x \xi_y v_F^2} I_1 +
\\ &&\hspace{-1.5cm}+\frac{2 \lambda \delta}{ N(\xi_x \xi_y)^2 v_F^3}
\left(I_1^2 \eta^2 + I_2^2-I_3 \right),
\end{eqnarray}
where the {\it in medium} integrals $I_0$, $I_1$, $I_2$ and $I_3$ are
respectively defined  by
\begin{eqnarray}\nonumber
I_0(\eta,\mu,T,{ \tilde{t}}) &=&\int  \frac{d^{2}{\bf p}}{(2\pi)^{2}}
\left [ \omega_p + T \ln \left( 1 + e^ {- \frac{\omega_p+\mu_t}{T}
  }\right) \right. \\ && \left. +T \ln \left( 1 + e^{-
    \frac{\omega_p-\mu_t}{T} }\right) \right] ,
\label{Integral0}
\end{eqnarray}
\begin{equation}
I_1(\eta,\mu,T,{ \tilde{t}})  =\frac{1}{2} \int \frac{d^{2}{\bf p}}{(2
  \pi)^{2}} \frac{1}{\omega_p}\left[1- n_p(\mu_t) -  {\overline
    n}_p(\mu_t) \right], 
\label{I1}
\end{equation}

\begin{equation}
I_2 (\eta,\mu,T,{ \tilde{t}})=  \frac{1}{2} \int \frac{d^{2}{\bf
    p}}{(2 \pi)^{2}} \left[ n_p(\mu_t) -  {\overline n}_p(\mu_t)
  \right],
\label{I2}
\end{equation}
and
\begin{widetext}
\begin{equation}
I_3(\eta,\mu,T,{ \tilde{t}} ) = \frac{1}{4}\int \frac{d^{2}{\bf p}}{(2
  \pi)^{2}} \frac{d^{2}{\bf q}}{(2 \pi)^{2}} {\bf p } \cdot {\bf q}~
\frac{1}{\omega_p \omega_q}\left[ n_p(\mu_t) +  {\overline n}_p(\mu_t)
  \right]\left[ n_q(\mu_t) +  {\overline n}_q(\mu_t) \right]\,.
\label{I3}
\end{equation}
\end{widetext}
In the above integrals the dispersion relation,  $\omega_p$, is given
by $\omega_p= \sqrt {{\bf p}^2 +\eta^2}$ while $\mu_t$ represents the
effective chemical potential defined by $\mu_t= \mu + |{\tilde {\bf
    t}}| |{\bf p}| \cos \theta$.  Note that vacuum terms have been
discarded in  $I_2$ and $I_3$ since in both cases the momentum
integrals turn out to be odd.  As usual, the {}Fermi-Dirac
distributions for particles and anti-particles, $n_p$ and $\bar{n}_p$,
respectively read
\begin{equation}
n_p(\mu_t) = \frac{1}{ e^ {(\omega_p - \mu_t)/T} +1},
\end{equation} 
and
\begin{equation}
\overline{n}_p(\mu_t) = \frac{1}{ e^ {(\omega_p + \mu_t)/T} +1}.
\end{equation}
Notice also that $I_2$ does not contribute at vanishing densities,
while $I_3$ does not contribute to isotropic matter and/or at
vanishing densities. As far as renormalizability is concerned let us recall that all integrals which depend on Fermi-Dirac distributions are regular (finite). So,  any potential ultra-violet divergences would be related to the  $\omega_p$-dependent (vacuum) terms present in Eqs.~(\ref{Integral0}) and (\ref{I1}). A well-established particularity of the  
$2+1\,D$ case is that the presence or absence of such divergences may depend on the adopted regularization scheme. If a sharp cut-off  is used to regularize both integrals the final results diverge when the cut-off is taken to infinity requiring the bare coupling to diverge~\cite{Klimenko92}.  On the other hand if one adopts the dimensional regularization scheme, as we do here, both integrals turn out to be finite implying that the bare coupling can also be taken as being finite\cite{Klimenko90}.    

\section{Optimization and gap equation beyond the large-$N$ limit}
\label{section4}

At order-$\delta$ the direct application of the optimization
criterion, Eq. (\ref{pms}), to the effective  potential  gives (see
also Refs.~\cite{Kneur:2007vj,Kneur:2007vm})
\begin{eqnarray}\label{PMSeq}
 && \left(  \bar{\eta}-\sigma +\frac{  \bar{\eta}  \lambda }{ N(\xi_x
    \xi_y) v_F}\bar{I}_1\right)\left(1 + \bar{\eta}\frac{\partial
  }{\partial \bar{\eta}} \right) \bar{I}_1 \nonumber \\ &&+ \frac{
    \lambda }{ 2N(\xi_x \xi_y) v_F} \left(  2 \bar{I}_2
  \frac{\partial \bar{I}_2}{\partial \bar{\eta}} - \frac{\partial
    \bar{I}_3}{\partial \bar{\eta} } \right)=0\,,
\end{eqnarray}
where the following relations have been used: $2 \eta I_1 - {\partial
  I_0}/{\partial \eta} = 0$, where $\bar{I}_i=I_i|_{\eta=\bar{\eta}}$
and ${\partial \bar{I}_i}/{\partial \bar{\eta}}={\partial
  I_i}/{\partial \eta}{|}_{\eta=\bar{\eta}}$. Note the extra term
$I_3$ in Eq.~(\ref{PMSeq}), which differs from previous
applications~\cite{Kneur:2007vj,Kneur:2007vm} due to the presence of a
nonvanishing tilt parameter.  It is important to remark that
Eq.~(\ref{PMSeq}) automatically leads to $\sigma = \bar{\eta}$ in the
limit $N \rightarrow \infty$, allowing us to exactly recover previous
large-$N$ results~\cite{Gomes:2021nem} for this model.  {}Finally, let
us emphasize that the PMS condition, Eq. (\ref{PMSeq}), has to be
solved in conjunction with the gap equation for ${\bar \sigma}$
\begin{equation}\label{gapeq}
{\bar \sigma}= \frac{4\lambda \bar{\eta}}{\xi_x \xi_y v_F} \bar{I}_1
\,,
\end{equation}
which follows from the stationary condition
\begin{equation}
\frac{ d V_{\rm eff}(\sigma)}{d \sigma} \Big{|}_{\bar{\sigma}} = 0\,.
\label{gap}
\end{equation}

In the next section, we present the numerical results obtained when
solving these coupled equations and study the resulting thermodynamics
of the model.

\section{Numerical Results}
\label{section5}

Let us start by recalling that, in $(2+1)$-dimensions, chiral symmetry
breaking only occurs when the GN coupling is negative\footnote{An
exception occurs when a magnetic field is present, see, e.g.,
Ref.~\cite{Kneur:2013cva} and references therein.}. Therefore, for
numerical purposes, it is convenient to replace $\lambda \to -
|\lambda|$ while defining the energy scale $\Lambda  = \pi \xi_x \xi_y
v_F/|\lambda|$  (not to be confused with a sharp momentum cut-off).  
{}For numerical convenience most quantities will be
presented in units of $\Lambda$ which, as we shall demonstrate,
corresponds to the gap energy predicted by the LN approximation at
$T=\mu=0$.  We are now in  position to investigate the possible
transition patterns associated with the four most representative
regions of the phase diagram.  

\subsection{The $T=0$ and $\mu=0$ case}
\label{section5A}

At vanishing temperatures and densities, one has $I_2=I_3 \equiv 0$,
hence, only $I_0$ and $I_1$ provide the vacuum contributions. As already emphasized, in
$(2+1)-$dimensions the latter integrals are finite when evaluated
within dimensional regularization. One then obtains
\begin{equation}
I_0 = -\frac{\eta^3}{6 \pi} \,, 
\label{I0T0}
\end{equation}
and 
\begin{equation}
I_1 = -\frac{\eta}{4 \pi}\,,    
\label{I1T0}
\end{equation}
in agreement with Refs.~\cite{Kneur:2007vj, Kneur:2007vm, Klimenko90}.
At $T=\mu=0$ the sole physical quantity of interest is the  chiral
order parameter, $\bar \sigma$. Then, we can use our new definitions,
together with Eq.~(\ref{I1T0}), to rewrite Eqs.~(\ref{PMSeq}) and
(\ref{gapeq}) as
\begin{equation}
 {\bar \eta} - {\bar \sigma} + \frac{{\bar \eta}^2}{4N\Lambda}  =0
\end{equation}
and
\begin{equation}
{\bar \sigma}  = \frac{{\bar \eta}^2}{\Lambda} \,.
\end{equation}
These relations yield
\begin{equation}
 {\bar \eta} = \frac{\Lambda}{\mathcal{F}(N)}  \,, 
\end{equation}
and 
\begin{equation}
 {\bar \sigma} = \frac{\Lambda}{\mathcal{F}(N)^2} \,,
 \label{sigbarT0}
\end{equation}
where we have defined the finite $N$ dependent term
\begin{equation}
\mathcal{F}(N) = 1-\frac{1}{4 N}\,.
\label{calF}
\end{equation}

As expected, this result exactly matches the one originally obtained
in Refs.~\cite{Kneur:2007vj,Kneur:2007vm}, where the planar GN model
was studied in the absence of a tilt in the Dirac cone. Equation~(\ref
{sigbarT0}) explicitly shows that the order parameter increases when
the number of fermionic species decreases. 

\subsection{The $T\ne 0$ and $\mu = 0$ case}

\begin{figure}[htb!]
    \centering    \includegraphics[scale=0.4]{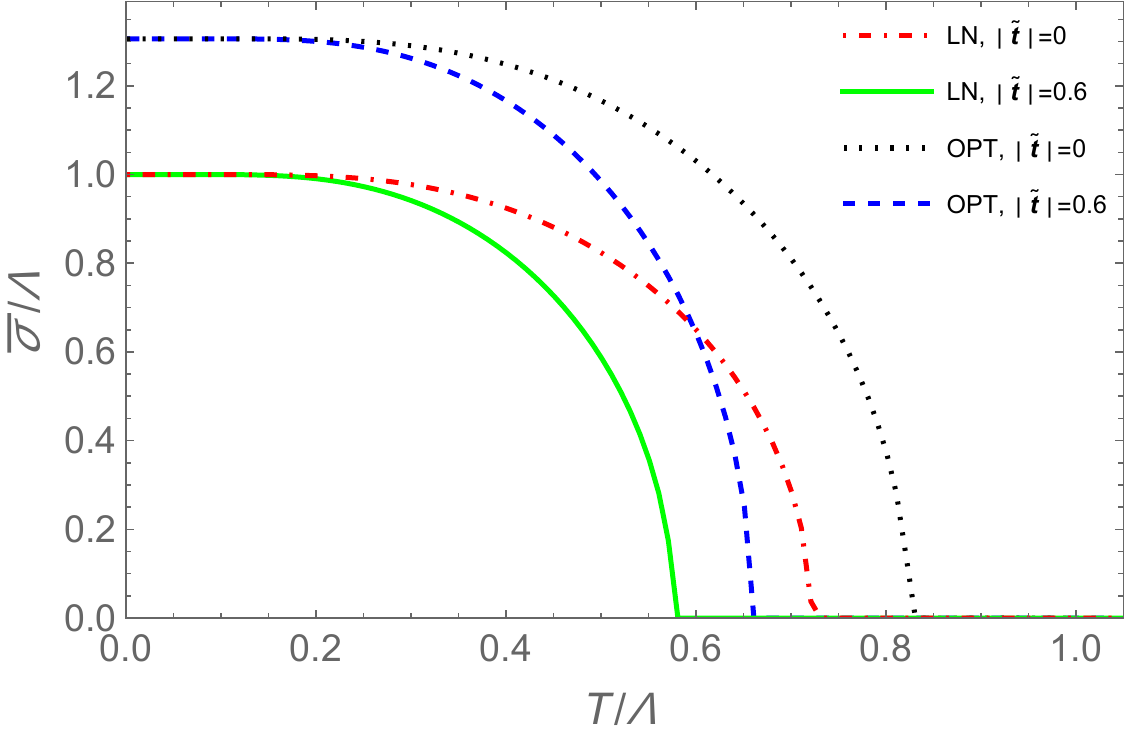}
    \caption{ The chiral order parameter, $\bar{\sigma}$, as a
      function of the temperature at $\mu=0$ and different values of
      the  tilt parameter, $|{\bf \tilde{t}}|$ (in units of
      $\Lambda$). Both approximations are compared considering  the
      case of $N=2$. }
    \label{fig1}
\end{figure}

Let us now describe the case of thermal matter at vanishing densities
by considering  the $T\ne 0$ and $\mu = 0$ case.  In this situation,
we still have $I_2=I_3\equiv 0$ such that the optimization condition
and the gap equation  respectively read
\begin{equation}
{\bar \eta} - {\bar \sigma}(T) - \frac{\pi {\bar \eta}}{N \Lambda}
I_1({\bar \eta},0,T,{\bf \tilde{t}}|)=0\,,
\label{PMSmuzero}
\end{equation}
and
\begin{equation}
{\bar \sigma}(T) = - \frac{4\pi}{\Lambda} {\bar \eta} I_1({\bar
  \eta},0,T,{\bf \tilde{t}}|).   
\label{gapmuzero}
\end{equation}
Now, substituting Eq.~(\ref {gapmuzero}) into Eq.~(\ref {PMSmuzero})
gives
\begin{equation}
1+ \frac{4\pi}{\Lambda} \mathcal{F}(N)  I_1({\bar \eta},0,T,{\tilde
  t})=0 \,,  
\label{newPMSmuzero}
\end{equation}
which can be solved numerically, for a fixed value of ${\bf \tilde{t}}|$, to
yield $\bar \eta (T)$. Then, the order parameter thermal behavior can
be directly obtained from Eq.~(\ref {gapmuzero}). {}Figure~\ref{fig1}
displays  ${\bar \sigma}(T)$ obtained at $N=2$ with the LN and the OPT
approximations. The figure  indicates that  the system undergoes a
second-order phase transition since the order parameter  smoothly
approaches  zero as $T\to T_c$. 

As one can also easily check, the thermal susceptibility, $\chi_T =
-d{\bar \sigma}/dT$, diverges at $T=T_c$. {}Finally, let us point out
that the critical temperature can be analytically evaluated as
follows. Recalling that ${\bar \sigma} (T_c)=0$ and that $\bar \sigma$
is proportional to $\bar \eta$, one can insert the analytical result
\begin{equation}
I_1(0,0,T,|{\bf \tilde{t}}|)= -\frac{2T \ln 2}{4 \pi \sqrt{1-t^2}} \,,
\end{equation}
into Eq.~(\ref{newPMSmuzero}) to obtain
\begin{equation}
T_c = \frac{\Lambda}{2 \ln 2 \mathcal{F}(N)} \sqrt{1-{
    \tilde{t}}^2}\,,
\label{tc}
\end{equation}
when $\mu=0$.

\begin{figure}[htb!]
    \centering    \includegraphics[scale=0.6]{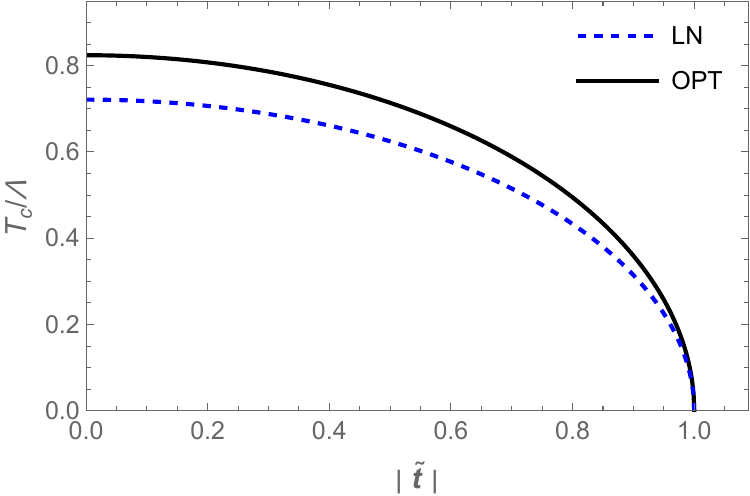}
    \caption{The critical temperature as a function of the tilt
      parameter,  $|{\bf \tilde{t}}|$,  at $\mu=0$ and $N=2$. The
      dashed line represents the LN result while the continuous line
      represents the OPT prediction. }
    \label{fig2}
\end{figure}

Our prediction for $T_c$, given by Eq. (\ref{tc}), reproduces the
result from Ref.~\cite{Kneur:2007vj} in the limit ${ \tilde{t}}
\rightarrow 0$ and the result from Ref.~\cite{Gomes:2021nem} in the
limit $N \rightarrow \infty$. The behavior of the critical temperature
as a function of the tilting parameter  is shown in {}Fig.~\ref{fig2}.

The results shown in this figure indicate that finite $N$ effects are
enhanced  at lower values of $|{\bf \tilde{t}}|$ when the difference between
the values of $T_c$ predicted by both approximations is larger. In
particular, the $T_c$ value predicted by the OPT is larger than the
one predicted by the LN approximation, which is expected since the
value of ${\bar \sigma}(0)$ predicted by the latter is also larger. As
$|{\bf \tilde{t}}|$ increases, this difference decreases such that both
approximations predict $T_c \to 0$ as $|{\bf \tilde{t}}|\to 1$ in accordance
with Eq.~(\ref {tc}). Therefore, the presence of a tilt in the Dirac
cone inhibits CSB. 

\subsection{The $T=0$ and $\mu \neq 0$ case}

Let us now consider the $T=0$ and $\mu \neq 0$ case.  Assuming a
positive and nonnull chemical potential, in the limit $T \rightarrow
0$, one finds $n_p(\mu_t) = \Theta(\mu_t - \omega_p)$ and
$\bar{n}_p(\mu_t) = \Theta(-\mu_t - \omega_p) =0$ where $\Theta(x)$
represents the Heaviside step function.  This result is valid for all
$\mu>0$, since $|\tilde{{\bf t}} \cdot {\bf p}| < \omega_p$.
Therefore, assuming $\eta>0$ and $\mu>0 $, one has 
\begin{equation}
I_0 =  - \frac{\eta^3}{6 \pi} + \left( \frac{\eta ^3}{6 \pi
}-\frac{\eta ^2 \tilde{\mu} }{4 \pi } +\frac{\tilde{\mu}^3}{12 \pi }
\right)~\Theta \left(\tilde{\mu}-\eta \right),
\end{equation}
where we have defined $\tilde{\mu} = {\mu}(1-|\tilde{\bf t}|^2)^{-1/2}$
while $p_F$ denotes the {}Fermi momentum. The latter  can be  obtained
from the inequality $\mu + |\tilde{\bf t}| p \cos \theta - \sqrt{p^2 +
  \eta^2}>0$. Namely,
\begin{equation}
p_F = \sqrt{\frac{\mu ^2-\eta ^2(1- |{\bf \tilde{t}}|^2 \cos^2 \theta
    )}{\left(1-|{\bf \tilde{t}}|^2\cos^2 \theta\right)^2}}+\frac{\mu
  |{\bf \tilde{t}}|\cos \theta}{1-|{\bf \tilde{t}}|^2 \cos^2 \theta} \,,
\end{equation}
from which the standard {}Fermi momentum, $p_F = \sqrt{\mu^2-\eta^2}$,
is recovered when  $|{\bf \tilde{t}}|=0$. {}Requiring $p_F \in \mathbb{R}$
implies that $\mu ^2-\eta ^2(1- |{\bf \tilde{t}}|^2 )> 0$, which, in turn,
produces the step function $ \Theta \left(\tilde{\mu}-\eta
\right)$. Likewise, for the integral  $I_1$, one obtains
\begin{eqnarray}
&&I_1   = - \frac{\eta}{4 \pi}-\frac{1}{4 \pi}\left(
  \tilde{\mu}-\eta\right)  \Theta \left(\tilde{\mu}-\eta \right)~~.
\end{eqnarray}
Note that $2 \eta I_1 = \frac{\partial I_0}{\partial \eta}$, as
expected.  Next, the integral $I_2$ can be written as
\begin{eqnarray}
 &&I_2  =\frac{1}{8 \pi \sqrt{1-|{\bf \tilde{t}}|^2}} \left(
  \tilde{\mu}^2-\eta^2 \right)  \Theta \left(\tilde{\mu}-\eta
  \right)~~.
\end{eqnarray}
To obtain $I_3$, one should first notice that the use of polar
coordinates, ${\bf p} = p( \cos \theta, \sin \theta)$, ${\bf q} = q(
\cos \theta', \sin \theta')$, leads to $\tilde{{\bf t}} \cdot {\bf p}
= |{\bf \tilde{t}}| p \cos \theta$  and $\tilde{{\bf t}} \cdot {\bf q} =
|{\bf \tilde{t}}| q \cos \theta'$. One then gets
\begin{eqnarray}
 && I_3 =  \frac{1}{64 \pi^2} \frac{|{\bf \tilde{t}}|^2}{1-|{\bf \tilde{t}}|^2}\left(
  \tilde{\mu}^2 - \eta^2 \right)^2  \Theta \left( \tilde{\mu}-\eta
  \right)\,
\end{eqnarray}
which explicitly shows that $I_3 = 0$ when $|{\bf \tilde{t}}| = 0$. Therefore,
the OPT effective  potential at first-order can be written as
\begin{eqnarray}
\!\!\!\!\!\frac{V_{\rm eff}(\mu)}{N} &=&\frac{-1}{\pi \xi_x \xi_y v_F^2}
\left \{\frac{\delta}{2} \sigma^2 \Lambda -   \frac{\eta^3}{3 } +
\delta {(\eta-\sigma)\eta^2}  + \frac{  \delta \eta^4}{8  N \Lambda}
\right .  \nonumber \\ &+\;&\!\!\!\!\!\!\left[\frac{\eta^3}{3 } -
  \frac{\eta ^2 \tilde{\mu} }{2  } +\frac{\tilde{\mu}^3}{6 } + \delta
  (\eta-\sigma)\eta( \tilde{\mu}- \eta) \right.
  \nonumber\\ &+\;&\!\!\!\!\!\!\left. \left .\frac{ \delta
    (\tilde{\mu}^2- \eta^2) \eta^2 }{8  N \Lambda} +\frac{  \delta
    \left( \tilde{\mu}^2-\eta^2 \right)^2}{32  N \Lambda}
  \right]\Theta \left(\tilde{\mu}-\eta \right) \right \} .  \nonumber
\\
\label{PotEffMUnulo2}
\end{eqnarray}

\begin{figure}[htb!]
    \centering    \includegraphics[scale=0.61]{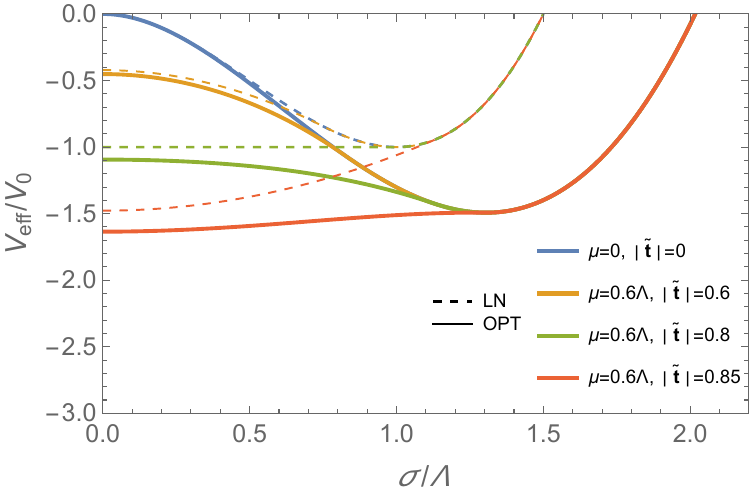}
    \caption{The effective thermodynamic potential at $T=0$ and
      $N=2$, for different values of $\mu$ and $|{\bf {\tilde t}}|$,
      as predicted by the LN approximation and the OPT (at first
      order). The normalization $V_0$ is given by
      $V_0/N=\Lambda^3/(6\pi \xi_x \xi_y v_F^2)$, as in Ref. \cite
      {Gomes:2021nem}.}
    \label{fig3}
\end{figure}

In {}Fig.~\ref{fig3} we show $V_{\rm eff}$, given by
Eq.~(\ref{PotEffMUnulo2}), as a function of $\sigma$ and for different
values of the tilt parameter. As one can notice, the figure displays a
typical first-order transition pattern. Namely, at $\mu=0$ the
effective potential has a maximum at $\sigma=0$ and a unique minimum
at  ${\bar \sigma} \ne 0$. As $\mu$ increases, the (unstable) maximum
at $\sigma=0$ becomes a point of inflection, marking the emergence of
the first spinodal. A slight increase of the chemical potential value
converts the inflection point, at $\sigma=0$, into a (metastable) {\it
  local} minimum, while a (stable) {\it global} minimum remains fixed
at ${\bar \sigma}\ne 0$. As expected,  the two non-degenerate minima
are separated by a barrier.  Then, when $\mu=\mu_c$, the two minima
become degenerate allowing for a first-order phase transition. When
$\mu$ is slightly higher than $\mu_c$, the minimum at the origin
becomes stable (global) while the one sitting at ${\bar \sigma} \ne 0$
becomes metastable (local). By further increasing the chemical
potential one eventually reaches the second spinodal, where the local
minimum at ${\bar \sigma} \ne 0$ becomes a point of inflection before
disappearing when $\mu$ further increases. Since the order parameter
is associated with the value of $\sigma$ at the (global) minimum, the
above discussion allows us to forecast that this quantity will suffer
a discontinuous transition from ${\bar \sigma} = \Lambda/
{\mathcal{F}(N)}^2 \to {\bar \sigma} =0 $ at the coexistence chemical
potential, $\mu=\mu_c$.

\begin{figure}[htb!]
    \centering    \includegraphics[scale=0.6]{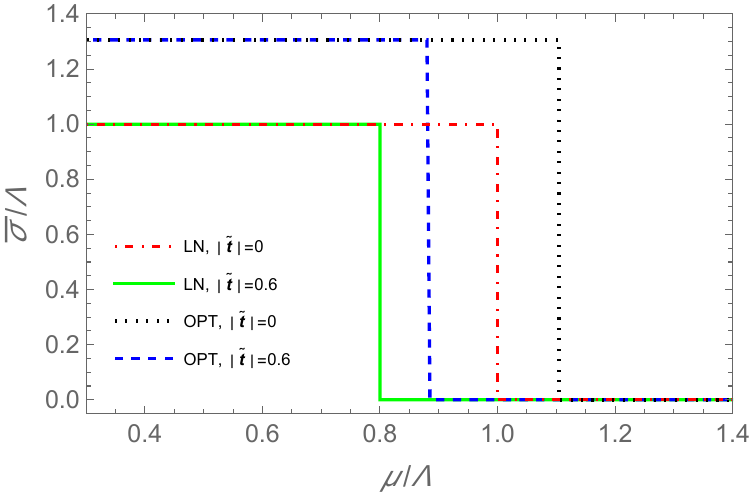}
    \caption{The chiral order parameter, $\bar \sigma$ (in units of
      $\Lambda$), as a function of $\mu/\Lambda$ for different values
      of the tilt parameter when $N=2$. In each case the values ${\bar
        \sigma}\ne 0$ and ${\bar \sigma}=0$ are linked by the
      (vertical) Maxwell line.}
    \label{fig4}
\end{figure}

In principle, to be able to analyze how the chiral parameter behaves
with $\mu$ one can proceed as in the  previously discussed case $(T\ne
0,\mu= 0)$ by considering the gap equation,
\begin{equation}
{\bar \sigma} = \frac{1}{\Lambda} \left [ {\bar \eta}^2 + {\bar
    \eta}({\tilde \mu} - {\bar \eta}) \Theta \left(\tilde{\mu}-\eta
  \right) \right ]\,.
\label{gap00}
\end{equation}
However, now the integrals $I_2$ and $I_3$ do not vanish and the
optimization equation becomes much more involving. As a matter of
fact, for certain values of  $\mu$ one may get three different $\bar
\eta$ solutions such that the gap equation will give three different
values for $\bar \sigma$, which ultimately  describe the three extrema
associated with the unstable, metastable and stable phases. In
practice, one can adopt a more pragmatic approach to obtain ${\bar
  \sigma}(\mu)$ by determining the {\it global} minimum of
$V_{\rm eff}(\mu)$ in a numerical fashion as we do here. The result is
shown in {}Fig.~\ref{fig4}.

\begin{figure}[htb!]
    \centering    \includegraphics[scale=0.6]{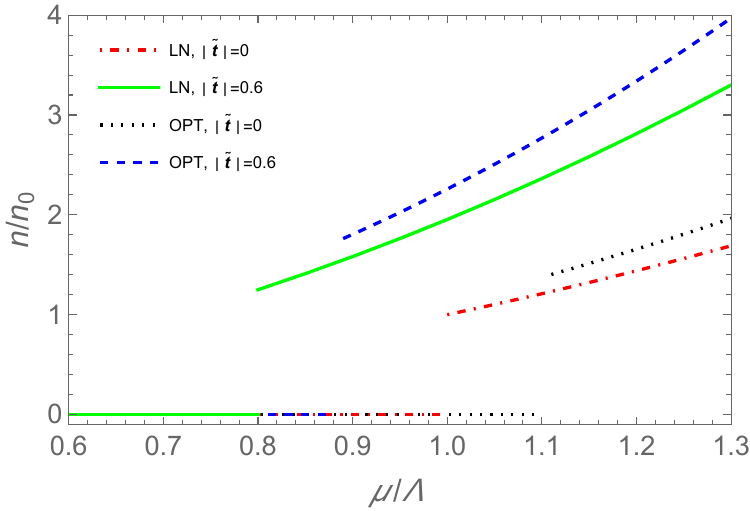}
    \caption{The fermionic number density, normalized by $n_0$ (see
      text), as a function of the chemical potential (in units of
      $\Lambda$) for $N=2$. Each curve indicates that a given  $\mu_c$
      the ``vacuum"  ($n=0$), which represents the insulator phase,
      coexists with  matter at a certain finite charge density
      (metallic phase). This coexistence phase could represent a
      semimetal.  }
    \label{fig5}
\end{figure}

As already mentioned, by evaluating the optimized effective  potential
at its minimum one obtains the optimized thermodynamic potential,
$\Omega = V_{\rm eff} ({\bar \sigma},{\bar \eta})$ which allows us to
obtain the net charge density, $n$. In the case of first order
transitions $n$ represents an interesting physical observable which
can also be viewed as an alternative order parameter. In a
thermodynamically consistent evaluation, the net charge number density
is written as 
\begin{equation}
n=-\frac{\partial \Omega}{\partial \mu} \,.
\end{equation}
Now, recalling that ${\bar \sigma}$ and ${\bar \eta}$ are
$\mu$-dependent and upon applying the chain rule one gets
\begin{equation}
\frac{\partial \Omega}{\partial \mu}= -n  + \frac{\partial {\bar
    \eta}}{\partial \mu}\frac{\partial \Omega}{\partial {\bar \eta}} +
\frac{\partial {\bar \sigma}}{\partial \mu}\frac{\partial
  \Omega}{\partial {\bar \sigma}} \,,
\label{omegaMU}
\end{equation}
which is thermodynamically  consistent since the optimization
criterion and the gap equation respectively require $\partial \Omega
/\partial {\bar \eta}=0$ and $\partial \Omega /\partial {\bar
  \sigma}=0$. More explicitly, one can write the number density {\it
  per fermionic specie} as
\begin{eqnarray}
\frac{n}{N} &=& \frac{1}{\pi \xi_x \xi_y v_F^2 \sqrt{1-|{\bf \tilde{t}}|^2}}
\left [ \frac{1}{2}({\tilde \mu}^2 - {\bar \eta}^2) + \delta {\bar
    \eta}({\bar \eta} - {\bar \sigma}) \right . \nonumber \\ &+& \left
  .\delta \frac{{\bar \eta}^2 {\tilde \mu}}{4N \Lambda}  +  \delta
  \frac{{\tilde \mu}({\tilde \mu}^2 - {\bar \eta}^2)}{8N \Lambda}
  \right ] \Theta \left(\tilde{\mu}-\eta \right)\,.
\label{density}
\end{eqnarray}
In {}Fig.~\ref{fig5} we illustrate the behavior of the charge number
density for different values of the tilt parameter as predicted by
both approximations considered in this work when $N=2$. Following
Ref. \cite {Gomes:2021nem} we have normalized $n/N$ by
$n_0/N=\Lambda^2/(2\pi \xi_x\xi_y v_F^2)$  which represents the LN
prediction at $t=0$. The figure clearly shows the coexistence of two
phases,  with $n=0$ (vacuum) and  $n\ne 0$ (charged), at the same
$\mu_c$. In principle, the phase with vanishing charge  density can
represent an insulator, while the phase with finite charge  density
can represent a metal. Within this scenario, the coexistence phase
could represent a semimetal.

The inspection of Eq.~(\ref{density}) reveals that the tilt, as well
as finite $N$ effects, favor higher density values as
{}Fig.~\ref{fig5} confirms. 

\begin{figure}[htb!]
    \centering    \includegraphics[scale=0.61]{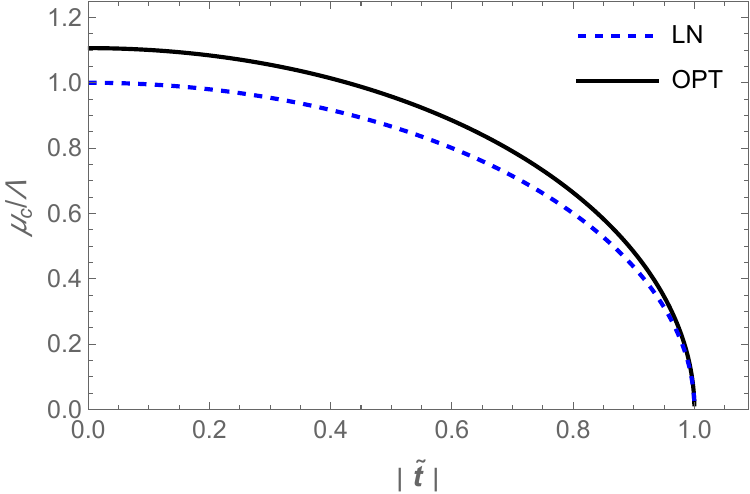}
    \caption{The coexistence chemical potential as a function of the
      tilt parameter,  $|{\bf \tilde{t}}|$,  at $T=0$ and $N=2$. The
      dashed line represents the LN result while the continuous line
      represents the OPT prediction. }
    \label{fig6}
\end{figure}

We can now attempt to find an equation to determine the numerical
value of the coexistence chemical potential. {}Following the reasoning
of Ref. \cite {Kneur:2007vm}, this can be achieved by  equating  the
values of the effective potential at the degenerate minima,
$V_{\rm eff}({\bar \sigma}\ne 0,  \mu_c <  {\bar \sigma})$ and
$V_{\rm eff}({\bar \sigma}= 0, \mu_c > {\bar \sigma} )$. This guarantees
that, despite having distinct densities, the two coexisting phases
share the same pressure (and temperature), as thermodynamics requires.
{}From section~\ref{section5A}, one can easily obtain the minimum
corresponding to the vacuum phase,
\begin{equation}
\frac{V_{\rm eff}({\bar \sigma}\ne 0,\mu_c <  {\bar \sigma})}{N} = -
\frac{1}{6 \pi \xi_x \xi_y v_F^2} \left ( \frac
     {\Lambda}{{\mathcal{F}(N)}}\right )^3 ,
\end{equation}
To obtain $V_{\rm eff}({\bar \sigma}= 0, \mu_c > {\bar \sigma})$, one can
use the gap equation~(\ref{gap00}) to set ${\bar \eta}={\bar \sigma}
\equiv 0$  in Eq.~(\ref{PotEffMUnulo2}). This yields the minimum
related to the charged phase
\begin{equation}
\frac{V_{\rm eff}({\bar \sigma}= 0,\mu_c > {\bar \sigma} )}{N} = -
\frac{\tilde{\mu}_c^3}{6 \pi \xi_y \xi_x v_F^2} \left(1 +\frac{3
  \tilde{\mu}_c }{16 N \Lambda }\right) \,,
\end{equation}
with the obvious notation $\tilde{\mu}_c =\mu_c (1-|\tilde{\bf
  t}|^2)^{-1/2}$. Therefore, the coexistence chemical potential
satisfies the equation
\begin{equation}
\left ( \frac {\Lambda}{{\mathcal{F}(N)}}\right )^3 - \tilde{\mu}_c^3
\left(1+\frac{3    \tilde{\mu}_c }{16  N \Lambda }\right)=0 .
\label{muc}
\end{equation}
Then, when $N=2$, Eq.~(\ref{muc})  sets the OPT prediction to  $\mu_c
\approx 1.106 \Lambda (1-|\tilde{\bf t}|^2)^{-1/2}$ while the LN
approximation predicts $\mu_c = \Lambda (1-|\tilde{\bf
  t}|^2)^{-1/2}$. In {}Fig.~\ref{fig6}, we  compare the results for
the coexistence chemical potential, obtained with both approximations,
as a function of the tilt parameter. The OPT predicts a higher value
for $\mu_c$ which, once again, could be expected owing to the fact
that ${\bar \sigma}(0)$ predicted by this approximation is larger. The
difference between the OPT and LN predictions decreases as the the
tilt increases since $\mu_c \to 0$ as $|{\bf \tilde{t}}|\to 1$.

\subsection{The $T \neq 0$ and $\mu \neq 0$ case}

\begin{figure}[htb!]
    \centering    \includegraphics[scale=0.4]{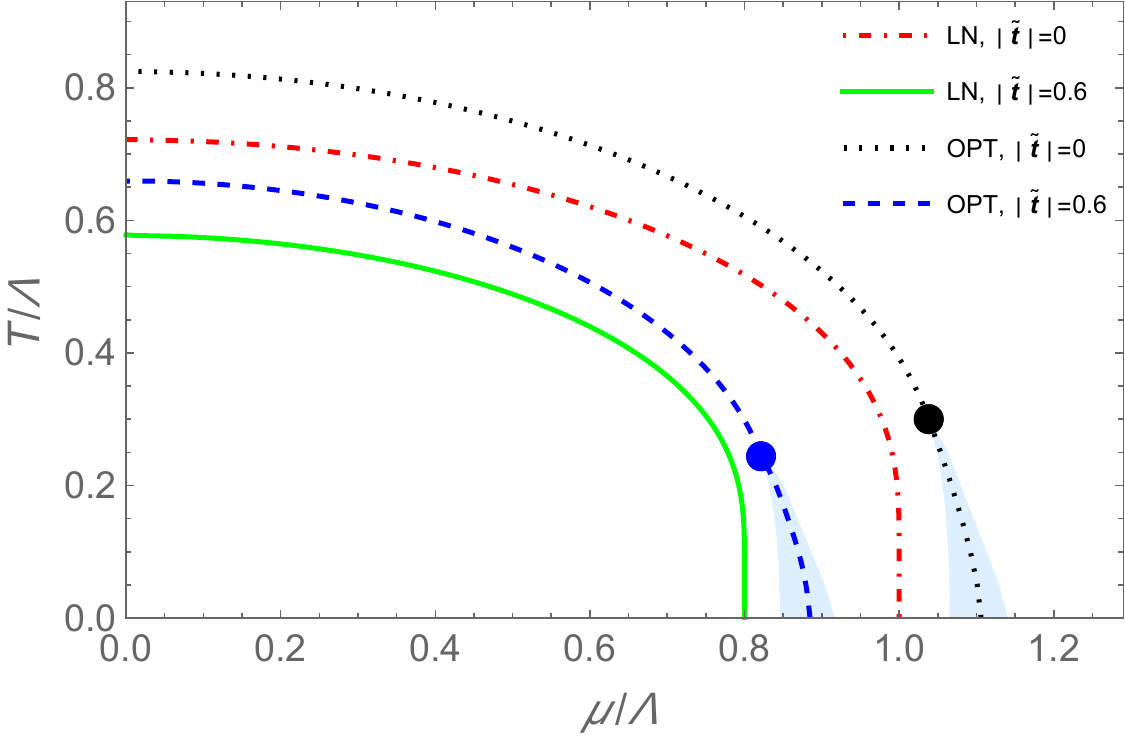}
    \caption{The phase diagram for $N$ = 2 on the $T-\mu$ plane. The
      OPT  tricritical points are indicated by the large dots. The
      second-order transition is represented by the lines above the
      tricritical points. The OPT predicts  that first-order
      transitions occur within the shadowed low-$T$ region below the
      tricritical points. This region is bounded by two spinodal lines
      while the central line marks the exact location  where the
      transition occurs. The LN approximation predicts a second-order
      phase transition at all finite temperatures and a first-order
      phase transition at $T=0$ only.  }
    \label{fig7}
\end{figure}

{}Finally considering   the case of finite temperatures  and densities
one needs to scrutinize the effective potential (or free energy) in
order to locate the regions  where first- and second-order phase
transitions occur. Then, one should be able to find the tricritical
point, located at ($T_{\rm tcr},\mu_{\rm tcr}$), which separates the
regions displaying distinct phase transitions.  Proceeding in a
numerical fashion one obtains the results displayed in
{}Fig.~\ref{fig7}. This figure portraits  the complete phase diagram
for the GN model with tilted Dirac cone  on the  plane spanned by  the
control parameters at hand ($T$ and $\mu$). The colored dots
associated to the OPT curves, for $|{\bf \tilde{t}}|=0$ and $|{\bf
  \tilde{t}}|=0.6$,  indicate the position of the respective
tricritical points. Above these points, the chiral transition  is of
the second kind. The shaded  areas that appear in the OPT results,
below the tricritical points, are associated with  first-order phase
transitions. Such  regions  are delimited by spinodal lines which
indicate the presence of  metastable phases. In between the two
spinodal lies a first-order transition line over which two stable
phases (with distinct densities) coexist at the same $T$ and $\mu$.

\begin{figure}[htb!]
    \centering    \includegraphics[scale=0.6]{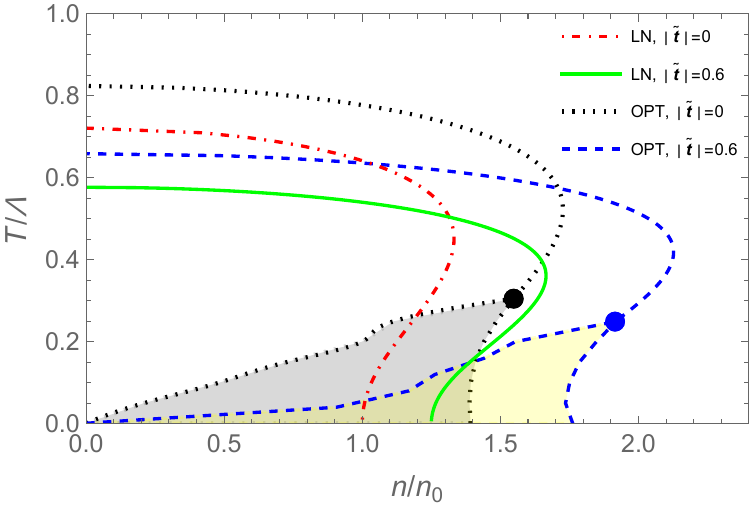}
    \caption{The phase diagram for $N$ = 2 on the $T-n/n_0$ plane. The
      OPT  tricritical points are indicated by the large dots. The
      second-order transition is represented by the lines above the
      tricritical points. The OPT predicts  that first-order
      transitions occur within the shadowed low-$T$ region below the
      tricritical points. This region is bounded by  binodal lines
      which, for a given temperature, indicate the two co-existing
      densities.  The LN approximation predicts a second-order phase
      transition at all finite temperatures and a first-order phase
      transition at $T=0$ only. }
    \label{fig8}
\end{figure}

To further analyze the nature of the phase transitions let us first
recall that, contrary to the chemical potential, the number density
does not represent a control parameter. Nevertheless, when combined
with the temperature this thermodynamical observable offers an
interesting  alternative in providing deeper insights into the
possible transition patterns, especially those related to
first-order phase transitions. This alternative framework is
represented by the phase diagram on the $T-n$ plane, as
Fig. \ref{fig8} displays. This figure shows that below the tricritical
temperature, the OPT predicts a first-order phase transition  region
which is bounded by binodal lines. One can also see that  for  a given
temperature, lower than $T_{\rm tcr}$, the binodal  indicates  the
existence of two distinct density values at which the different phases
(symmetric and nonsymmetric)  coexist. The inner region is associated
to the existence of a  mixed phase, characterized by metastable and
unstable states. Recalling that  within the present model, the typical
gap energy (at $T=\mu=0$) has a value close  to $\Lambda$ the OPT
therefore predicts the existence of a mixed phase from $T=0$ to
roughly $T \sim \Lambda/4$. The possible existence of such a mixed
phase, missed by the LN (and mean field) approximation, represents our
main result. 

{}For completeness, in Table~\ref{tab1}, we show the predictions for
the tricritical control parameters, $T_{\rm tcr}$   and  $\mu_{\rm
  trc}$, as well as for the charge density at this location, $n_{\rm
  trc}= n (T_{\rm tcr},\mu_{\rm tcr})$, considering some
representative values of the tilt parameter. The numerical values of
$T_{\rm tcr}$ and $\mu_{\rm trc}$ can be obtained in different
ways. Here, we have numerically determined the values at which the two
degenerate minima, associated with the first order transition, merge
and become the unique minimum (at $\bar \sigma =0$),  characterizing a
phase transition of the second kind. A possible alternative to
determine $T_{\rm tcr}$   and  $\mu_{\rm trc}$  is to apply Landau's
expansion to the  free energy ($V_{\rm eff}$),  following the prescription
described in Ref.~\cite{Kneur:2007vj} (where only the untilted case
has been considered).

{}The results shown in Table~\ref{tab1} indicate that the values of
the tricritical control parameters, $T_{\rm tcr}$ and  $\mu_{\rm
  trc}$, decrease as $|{\bf \tilde{t}}|$ increases. This result could be
anticipated since the tilt enhances the shrinkage of the region with
broken chiral symmetry, as we have already discussed. On the other
hand, the observable $n_{\rm tcr}$ assumes higher values as the tilt
increases.

\begin{table}[htb!]
\caption{Values for the tricritical temperature, tricritical chemical
  potential, and charge density (at the tricritical point) for
  different values of the tilting parameter,  as predicted by both
  approximations when $N=2$.}
\begin{tabular}{ c|c|c|c|c }
 \hline Method & $|{\bf \tilde{t}}|$	& $T_{\rm tcr}/\Lambda$ &
 $\mu_{\rm trc}/\Lambda$ & $n_{\rm tcr}/n_0$ \\ \hline & 0 & 0& 1.0 &
 1.0\\ & 0.3 &0 &  0.954& 1.048\\ LN	& 0.6 &0	&  0.800&
 1.251\\ & 0.9 &0	& 0.436& 2.299\\ \hline & 0 & 0.305& 1.040 &
 1.557\\ & 0.3 & 0.285 & 0.990 & 1.613\\ OPT	& 0.6 &	0.245 & 0.820
 & 1.902 \\ & 0.9 &	0.135 & 0.455 & 3.617\\ \hline
 \end{tabular}
\label{tab1}
 \end{table}

\subsection{Applications}

It should be noticed that Weyl like materials are still rather rare.
The difficulty to find 2DWSM lies in the fact that a  two-dimensional fermionic system does not have topological protection against gap formation in comparison to the 3D counterpart. Despite of this, some materials are known, as for instance,  quinoid-type graphene and $\alpha$-(BEDT-TTF)$_2 I_3$ organic conductors~\cite{goerbig2008,Hirata} and, more recently, the cases of Epitaxial Bismuthene~\cite{2303.02971} and porous Si/Ge structures that have been proposed as a 2DWSM~\cite{2305.05756}. Several candidates for bidimensional Dirac materials have also been proposed, such as  Si and Ge honeycomb lattices as well as porous and crystalline organic structures among a myriad of other possibilities (for a recent review,
see, e.g., Ref.~\cite{runyu}.  Keeping that in mind let us  discuss some implications of our results, based on some of the available experimental data for some of the existing bi-dimensional materials.

We can now use a phenomenological input in order to fix the GN scale,
$\Lambda$,  so that our theoretical predictions can be ultimately
compared to some of the available experimental results. The aim of
such exercise is just to check whether the OPT predictions lie within
the ball park range of the available experimental data.  As a starting
point, we can consider that current experiments are carried out at
room temperature ($300\, K$) or close to it. We will also suppose that
these temperatures lie within the region of the second-order phase
transition and just above the (highest) tricritical temperature (see
{}Fig.~\ref{fig8} for reference). 

According to  the results shown in Table~\ref{tab1}, the highest value
for the tricritical temperature is  $T_{\rm trc} = 0.305 \, \Lambda$,
which occurs for $|{\bf \tilde{t}}|=0$.  We can then assume that the
room temperature ($300\,K$) lies slightly above $T_{\rm trc}$, but
still in the second-order transition region and take for example
$T_{\rm exp}= 0.35 \, \Lambda = 0.0258\, {\rm eV}>T_{\rm trc} $. This then
allows us to fix the GN scale to be $\Lambda_{\rm exp} = 0.0738 \, {\rm
  eV}$. Note that, since $\Lambda \propto 1/|\lambda|$, this is
basically the equivalent of fixing the coupling, which represents the
only energy scale present in the $(2+1)-$dimensions GN
model. Considering this value of $\Lambda$ as the numerical scale used
to produce Fig. \ref{fig7}  one gets, at $T=T_{\rm exp}$, the values
$\mu_{\rm exp} \approx 1.02  \Lambda_{\rm exp} = 0.0753 \, {\rm eV}$ and
${\bar \sigma}_{\rm exp} \approx 1.3 \Lambda_{\rm exp} = 0.096 \, {\rm eV}$
for the chemical potential and gap energy respectively. As discussed
in Ref. \cite{graph}, both values lie within  the bounds of
phenomenological  predictions.  {}Finally, one can check the actual
charge density number at $T_{\rm exp}$ by taking $n/n_0 \approx 1.6$  (see
{}Fig.~\ref{fig7} and  Table~\ref{tab1}). Then, using
\begin{equation}
n_{\rm exp} = \frac {N}{2\pi \xi_x \xi_y v_F^2} \left ( \frac{n}{n_0}
\right ) \Lambda_{\rm exp}^2, 
\end{equation}
at $N=2$, $\xi_x=\xi_y\equiv 1$, and $v_F/c=1/300$ one gets $n = 1.62
\times 10^{10} {\rm cm}^{-2}$ which is very similar to the result $n =
1.2 \times 10^{10} {\rm cm}^{-2}$ obtained for instance in experiments
devoted to measure the Casimir interaction between a Au-coated sphere
and a graphene-coated ${\rm SiO}_2$ layer on the top of a Si
plate~\cite{graph}.  To reach the first-order transition region (with
its mixed phase) one should further lower the temperature. {}For
example, at $|{\bf \tilde{t}}| =0$ one may set  $T= 0.3 \Lambda
= 0.0221\, {\rm eV}<T_{\rm trc}$, which corresponds to $255.29 \,K$
whereas, at $|{\bf \tilde{t}}| =0.6$, one can consider $T= 0.235
\Lambda = 0.0173\, {\rm eV}<T_{\rm trc}$, which corresponds to $200
\,K$. 

By going to extremely low temperatures, one eventually will reach a
region  where two phases, with very different densities, can
coexist. {}For example, when $|{\bf \tilde{t}}| =0$ and $T= 0.05
\Lambda = \, 0.037 {\rm eV}$ ($42.82\,K$)  one can have the low
density ($n = 0.2125 \times 10^{10} \, {\rm cm}^{-2}$) gapped
(nonsymmetric) phase  coexisting with the high density ($n = 1.405
\times 10^{10}\, {\rm cm}^{-2}$)  nongapped (symmetric) phase. 

{}Furthermore, recent numerical work on Weyl semimetals~\cite{crippa}
have shown a first type topological quantum phase transition as a function of the
interaction. It would be interesting to look at this type of model in regimes 
where a similar transition, as a function of the chemical potential and temperature,
can appear,  as predicted by our results.

{}Finally, one should also note that from some of the available experimental results on (quasi-)planar materials (such as for example in Ref. [54]), the linear simple Weyl cone structure is well described for  energies below and around ${\cal O}(10 {\rm meV})$. {}At appropriate values of the energy scale ($\Lambda$), and  for corresponding  temperature and chemical potential not much higher than $\Lambda$, the possible nonlinearities in the dispersion relation are expected to remain small and our results should provide a reasonable qualitative description of the real physical systems. 
In particular, the estimates provided above, fall in the ballpark 
of what one would expect for the linear (relativistic) dispersion  
to be a reasonable approximation.


\section{Conclusions}
\label{section6}

Considering the tilted $(2+1)$-dimensional GN model, we have analyzed
the phase transition patterns associated with the breaking or restoration
of chiral symmetry within planar Weyl type of materials. In order to
incorporate  finite-$N$ effects, we have employed the OPT
approximation to evaluate the effective thermodynamic potential
(Landau's free energy) to the first non-trivial order. This strategy
has allowed us to consider a two-loop contribution (of order $1/N$) to
improve over existing large-$N$ results~\cite {Gomes:2021nem}. 

Starting with the simplest ($T=\mu=0$) case, where the tilt does not
contribute, we have reproduced previous OPT results confirming that
finite $N$ effects produce chiral order parameter values that are
higher  than the ones predicted at  large-$N$.  Proceeding to the case
of thermal matter at vanishing densities, we have shown that the
system undergoes a second-order phase transition at a critical
temperature given by $T_c = \Lambda (1-|{\bf \tilde{t}}|^2)^{1/2}
\,[2 \ln 2 \mathcal{F}(N)]^{-1}$. This result suggests that finite $N$
corrections lead to phase transitions taking place at   higher $T_c$
values than those predicted by the LN approximation. On the other
hand, the tilt (parametrized by $|{\bf \tilde{t}}|$) favors a more
disordered phase such  that $T_c \to 0$ as $|{\bf \tilde{t}}| \to 1$.
Considering the case of a finite chemical potential at zero
temperature, we have once again reproduced previous predictions
confirming that chiral symmetry is restored through a first-order
phase transition. The coexistence chemical potential, $\mu_c$, at
which the transition takes place, increases with decreasing values of
$N$. Also, in this case, we observe that the tilt enhances the
restoration of chiral symmetry since  $\mu_c \to 0$ as $|{\bf
  \tilde{t}}|\to 1$. In summary,   the phase transition patterns
predicted by the LN and by the OPT for these three particular cases
are {\it qualitatively} the same. {}From a more quantitative
perspective, the OPT predictions indicate that finite $N$ effects
favor the ordered phase. 

Our investigation  shows that the alluded
  qualitative agreement  ceases to be observed at  finite temperature
  and chemical potential, when the phase transition boundaries
  predicted by both approximations become very different. As it is
  well established, the LN phase diagram displays a second-order
  transition line for all $T>0$, since a first-order transition is
  predicted to occur only at $T=0$. However, we have explicitly shown
  that this situation turns out to be   an artifact of  the LN
  approximation, and already at the first non-trivial order, the OPT
  is capable of predicting a phase diagram which is more in line with
  what continuity arguments~\cite{Kogut:1999um} suggest. More
  specifically, finite $N$ effects produce a  first-order transition
  line  that starts at $T=0$ and terminates at a tricritical point
  roughly located at a (tricritical) temperature whose value is about
  $20\%$ to $30\%$ of the value of the {\it ``in vacuum"} gap energy
  ($\sim \Lambda$). Physically, one of the most important consequences
  of the  first-order transition line predicted here is that it allows
  for two distinct phases of a Weyl type of material to co-exist at
  low  temperatures. Therefore,  a Weyl-type of planar materials should
  have a  phase transition structure much richer than one predicted by
  the LN (or mean-field) approximation. Here, the prediction of such
  a mixed phase has been entirely derived upon using quantum field
  theory methods.  Possible further extensions of the present
  application might be for example to extend the LN study of
  superconducting phase transitions in planar fermionic models with
  Dirac cone tilting~\cite {Gomes:2022dmf} such as to include finite
  $N$ effects and also looking for the effects of magnetic fields when
  applied to the system.   

\begin{acknowledgements}

Y.M.P.G. is supported by a postdoctoral grant from Funda\c{c}\~ao
Carlos Chagas Filho de Amparo \`a Pesquisa do Estado do Rio de Janeiro
(FAPERJ), grant No. E-26/201.937/2020. E.M. is supported by a PhD grant from Conselho Nacional de
Desenvolvimento Cient\'{\i}fico e Tecnol\'{o}gico (CNPq).  
M.B.P. is partially supported by Conselho Nacional de
Desenvolvimento Cient\'{\i}fico e Tecnol\'{o}gico (CNPq), 
Grant No 307261/2021-2. 
M.B.P. and R.O.R. also achnowledge support from 
Coordena\c{c}\~ao de Aperfei\c{c}oamento de Pessoal de N\'{\i}vel 
Superior - Brasil (CAPES) - Finance Code 001.  R.O.R. is also partially supported by
research grants from Conselho Nacional de Desenvolvimento
Cient\'{\i}fico e Tecnol\'ogico (CNPq), Grant No. 307286/2021-5, and
from {}Funda\c{c}\~ao Carlos Chagas Filho de Amparo \`a Pesquisa do
Estado do Rio de Janeiro (FAPERJ), Grant No. E-26/201.150/2021. This
work has also been financed  in  part  by Instituto  Nacional  de
Ci\^encia  e Tecnologia de F\'{\i}sica Nuclear e Aplica\c c\~{o}es
(INCT-FNA), Process No.  464898/2014-5.  R.O.R. would like to thank
the hospitality of the Department of Physics McGill University and
where this project was started.

\end{acknowledgements}



\end{document}